\documentclass[lettersize,journal]{IEEEtran}
\usepackage{amsmath,amsfonts}
\usepackage{algorithmic}
\usepackage{algorithm}
\usepackage{array}
\usepackage[caption=false,font=normalsize,labelfont=sf,textfont=sf]{subfig}
\usepackage{textcomp}
\usepackage{stfloats}
\usepackage{url}
\usepackage{verbatim}
\usepackage{graphicx}
\usepackage{cite}
\usepackage{multirow}

\begin{document}

\title{From Lagging to Leading: Validating Hard Braking Events as High-Density Indicators of Segment Crash Risk}

\author{Yechen Li$^a$, Shantanu Shahane$^a$, Shoshana Vasserman$^b$, Carolina Osorio$^c$, Yi-fan Chen$^a$, Ivan Kuznetsov$^a$, Kristin White$^a$, Justyna Swiatkowska$^a$, Neha Arora$^a$, Feng Guo$^d$

\bigskip

$^a$Google Research, Mountain View, CA, USA

$^b$Stanford Graduate School of Business, CA, USA

$^c$HEC Montr\'{e}al, Montr\'{e}al, QC, Canada

$^d$Virginia Polytechnic Institute and State University, VA, USA
}

\date{December 9, 2025}



\maketitle

\begin{abstract}


Identifying high crash risk road segments and accurately predicting crash incidence is fundamental to implementing effective safety countermeasures. While collision data inherently reflects risk, the infrequency and inconsistent reporting of crashes present a major challenge to robust risk prediction models. The proliferation of connected vehicle technology offers a promising avenue to leverage high-density safety metrics for enhanced crash forecasting. A Hard-Braking Event (HBE), interpreted as an evasive maneuver, functions as a potent proxy for elevated driving risk due to its demonstrable correlation with underlying crash causal factors. Crucially, HBE data is significantly more readily available across the entire road network than conventional collision records. This study systematically evaluated the correlation at individual road segment level between police-reported collisions and aggregated and anonymized HBEs identified via the Google Android Auto platform, utilizing datasets from California and Virginia. Empirical evidence revealed that HBEs occur at a rate magnitudes higher than traffic crashes. Employing the state-of-the-practice Negative-Binomial regression models, the analysis established a statistically significant positive correlation between the HBE rate and the crash rate: road segments exhibiting a higher frequency of HBEs were consistently associated with a greater incidence of crashes. This sophisticated model incorporated and controlled for various confounding factors, including road type, speed profile, proximity to ramps, and road segment slope. The HBEs derived from connected vehicle technology thus provide a scalable, high-density safety surrogate metric for network-wide traffic safety assessment, with the potential to optimize safer routing recommendations and inform the strategic deployment of active safety countermeasures.
\end{abstract}

\begin{IEEEkeywords}
Hard Braking Event, Traffic Safety, Crash Prediction, High Risk Road Segment
\end{IEEEkeywords}

\section{Introduction}
\IEEEPARstart{T}{raffic} crash statistics have been considered the ‘gold standard’ in traffic safety evaluation, as they are directly correlated to fatality, injury, property damage, and other related factors [1]. However, there are limitations in crash data: (1) Crash data is limited. For  local roads, it could take years to observe sufficient number of crashes for reliable safety evaluation; (2) Crash data can be difficult to collect, and the data collection standards are inconsistent across organizations and regions. (3) The retrospective nature of crash data creates a reporting lag that hinders proactive, real-time safety interventions.  Crash surrogate data such as  near-crashes [2], HBEs, and other safety performance metrics, have been widely used in safety research. The ubiquity of smartphones and mobile map applications, provide an opportunity to assess traffic safety at the global road network level. A key issue for using smartphone-based metrics, as for all non-crash safety events, is whether they indicate driving risk and how that translates to crash risk.

For decades, the traffic safety research community has utilized non-crash event data to evaluate safety. In the 1980s the Federal Highway Administration published guidelines for evaluating safety using the ‘traffic conflict technique’, which is based on manually observed conflicts between vehicles and other road users [3]. The rationale was that traffic conflicts occur at a much higher frequency than crashes, and locations with high traffic conflict risk are more prone to crashes. Following similar logic, more sophisticated surrogate metrics have been developed based on advancement in sensor technologies and analytic methods including: time-to-collision, post encroachment time, time advantage, gap time, and braking time ([4], [5], [6], [7]). These proximity-based surrogates quantify the spatial separation or temporal interval between vehicles or objects during an impending collision course. They have been shown to reflect hazardous and crash prone scenarios associated with a specific road segment and are typically based on sensors and cameras at fixed locations. 

Non-crash safety surrogate data are also important for automated driving systems (ADS).    A recent study indicated that decades of data will be needed to confirm the safety of ADS based on crashes [8]. A subsequent report presented a framework to use leading measures, “i.e., proxy measures of driving behaviors correlated to safety outcomes”, to assess the safety of ADS without relying solely on crashes [9]. The concept of connected vehicles facilitates the comprehensive sharing of vehicle and driving data across different platforms. The connected vehicle data offers an avenue for identifying various leading measures obtainable through onboard vehicle sensors. Leading measures based on vehicle sensors have been demonstrated and implemented through large-scale naturalistic driving studies. For example, near-crashes, safety-critical conflicts, high G-force events, and unintentional lane deviation events have been used to evaluate driver distraction, fatigue, and teenage driver risk ([10], [11], [12], [13]).   

One central issue for safety surrogate measures is the validity, meaning that there is question whether the surrogate metric truly represents crash risk and adequately supports research objectives. Researchers have debated whether surrogates and crashes are different by definition, and validity hinges on whether surrogate and crash-based conclusions align, thus depending on objectives. [14]. For example, Guo et al [11] proposed two principles to validate whether near-crashes can be used to evaluate driver distraction: (1) The causal mechanism should be similar; (2) there should be a strong frequency correlation between the two types of events. Understanding the causal mechanisms of varying types of crashes necessitates expert domain knowledge and detailed crash investigation analysis for validation.  Strong correlations as evidenced by the significant association via statistical models has been the primary method for examining the validity of surrogates ([12], [11], [15]). 

To monitor and log the surrogate events, vehicles today come equipped with various sensors like LiDAR, camera, radar and others that record real-time driving information, such as speed, acceleration, deceleration, and occurrences of hard braking. This research leverages sensor data from connected vehicles, specifically focusing on the speed and acceleration shifts observed during a hard braking incident. The HBEs are also predictable to support safety-aware navigation ([16], [17]). This study addresses the safety premise more directly by evaluating the relationship between HBEs 
and state-collected police-reportable crashes for California and Virginia, with consideration for other risk factors such as road infrastructure.

The research leverages the most current and effective methodologies in traffic safety research, specifically focusing on the rigorous evaluation of risk factors. The analytical framework adheres to the comprehensive guidelines and recommended approaches outlined in the Highway Safety Manual (HSM [1]). This framework is crucial for accurately assessing the correlation between hazardous driving behaviors (HBEs) and the incidence of crashes. To achieve this, the study utilizes a combination of publicly available crash databases and proprietary HBE datasets. The specific objectives of this study are multifaceted and encompass the following key areas:
(i) To evaluate the association between HBEs and crash rate under the Highway Safety Manual framework; 
(ii) To evaluate how this association is affected by other road characteristics.

\section{Data and Method}
\subsection{Data}
This study integrates publicly available collision records with proprietary safety metrics and road segment characteristics. The publicly available crashes for the states of Virginia (10 years) and California (10 years) are downloaded from virginiaroads.org and https://data.ca.gov/dataset/ccrs, respectively. To accurately account for exposure and risk, we extracted aggregated and anonymized HBE measurements and estimated traffic volumes from Google Maps Traffic Trends. These data sources are aligned at road segment level. 


The HBEs are based on vehicle wheel speed during active navigation sessions. The change in speed, i.e., longitudinal acceleration, is derived and used for HBE identification. An HBE is defined as an instance where  longitudinal deceleration exceeds a predefined threshold.  To reflect HBE risk, the HBE counts are normalized by the distance driven by the vehicles with active HBE detection. HBE data is aggregated and anonymized using differential privacy preserving methods.






In addition to dynamic vehicle metrics, the study extracts road attributes from Google Maps road network data to control for environmental factors.  Key infrastructure characteristics that can potentially impact safety are also extracted, including road types, presence of a ramp on the road segment, and the changes in the number of lanes etc.  Utilizing infrastructure data from Google Maps, road segments were categorized into a four-tier classification system for modeling purposes: Type 1 represents local roads or rural routes serving residential areas; Type 2 comprises arterial roads facilitating moderate- to high-capacity transit between neighborhoods and commercial hubs; Type 3 includes non-controlled access highways typically characterized by level-grade intersections; and Type 4 encompasses controlled-access highways featuring grade-separated crossings and interchanges.

Road segments missing critical information, such as length and traffic volume information, were omitted from the analysis 

\subsection{Model structure}
The analysis follows the Highway Safety Manual approach (HSM, National Research Council 2010) using Poisson/Negative Binomial (NB) Generalized Linear Models. As preliminary analysis indicates the presence of over-dispersion, the NB model is used per the recommendation of HSM.  The formulation of the NB model is described as follows.   Denote $Y(s)$ as the number of crashes on road segment $s$. The NB models the total crashes as

$Y(s) \sim NB(V_s * L_s * \lambda(c_s), \kappa)$

where: $V_s$ is the traffic volume for segment $s$;  $L_s$ is the length of the segment;  $c_s$ is the expected crash rate for segment $s$.  The term $V_s * L_s$ is the vehicle-distance measure of exposure. The exposure reflects that the expected number of crashes are higher on longer segments or busier segments. For any given segment, the exposure was calculated as $L_s * V_s * m$, where $m$ is the number of years where crashes were observed. This exposure measure reflects the vehicle-miles traveled over the years where crashes were observed. This is an over-disperson parameter. 

The expected crash rate ($c_s$) is a function of the characteristics of the segment $s$. The standard GLM models the expected crash rate as
$\lambda(c_s) = \exp(\beta_0 + \beta_1 X_{1,s} + \dots + \beta_p X_{p,s})$
where $\beta$'s are regression coefficients and $X$'s are the characteristics associated with segment $s$.  This study focused on the association between HBEs and crashes and how the association varied by road type and various infrastructure characteristics.

\section{Results}
\subsection{Exploratory data analysis}
The analysis datasets include road characteristics, HBEs, and crashes from the states of California and Virginia. Summary statistics of the road segments are shown in Table 1 The analysis was restricted to road segments meeting specific criteria: those with a valid crash rate, valid rates for HBE and belonging to one of the four primary road classifications.

\begin{figure}[!t]
\centering
\includegraphics[width=2.5in]{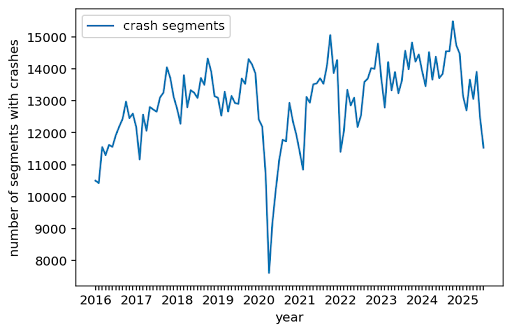}
\caption{Number of road segments with crashes every month}
\label{fig_1}
\end{figure}
\begin{figure}[!t]
\centering
\includegraphics[width=2.5in]{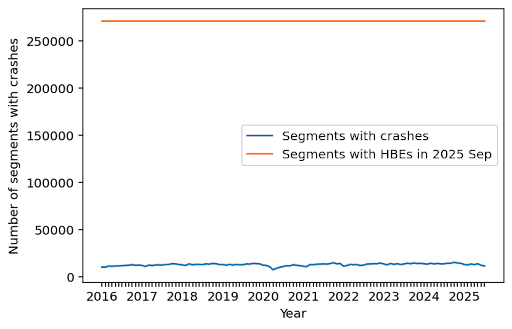}
\caption{There are 18x more road segments with HBEs than with crash data.}
\label{fig_2}
\end{figure}

Figure 1 displays the seasonal variations in crashes over the last decade. The Figure 2 illustrates that the number of road segments experiencing observed HBEs is 18 times greater than those with reported crashes. Note that the HBE data consists of data for September 2025, while the crash data is monthly over a multi-year period.

\begin{table*}[t]
\centering
\resizebox{405pt}{35pt}{
\begin{tabular}{|l|c|c|c|c|c|c|c|c|c|c|}
\hline
\multirow{2}{*}{\text{Statistical Metrics}} & \multicolumn{2}{c|}{\textbf{No of Segments}} & \multicolumn{2}{c|}{\textbf{Total length}} & \multicolumn{2}{c|}{\textbf{Number of crashes}} & \multicolumn{2}{c|}{\textbf{Average crash rate}} & \multicolumn{2}{c|}{\textbf{Average HBE rate}} \\ 
\cline{2-11} 
 & \textbf{CA} & \textbf{VA} & \textbf{CA} & \textbf{VA} & \textbf{CA} & \textbf{VA} & \textbf{CA} & \textbf{VA} & \textbf{CA} & \textbf{VA} \\ 
\hline
\textbf{Type 1 (Local roads)} & 48,072 & 12,421 & 18,264 & 5,409 & 201,656 & 46,897 & 7.54 & 6.82 & 0.03 & 0.02 \\ 
\hline
\textbf{Type 2 (Arterial roads)} & 12,510 & 13,322 & 4,089 & 6,344 & 62,197 & 65,607 & 7.03 & 8.14 & 0.03 & 0.03 \\ 
\hline
\textbf{Type 3 (Non-controlled) access highways} & 16,768 & 25,045 & 9712 & 13,483 & 99,239 & 143,903 & 6.20 & 6.44 & 0.02 & 0.002 \\
\hline
\textbf{Type 4 (Controlled) access highways)} & 22,752 & 14,730 & 15,266 & 10,669 & 945,967 & 171,639 & 5.07 & 2.11 & 0.003 & 0.003 \\
\hline
\end{tabular}%
}
\caption{Comparison of Road Statistics between CA and VA}
\label{tab:road_stats}
\end{table*}

Table I shows the statistics of road segments for California and Virginia, after data anonymization. Crash rate is measured as crashes per million vehicle miles traveled (i.e.,  number of crashes/(AADT*365* years*length of the segment)/1000,000 ). The HBE rate is measured as HBEs per vehicle miles traveled. The year-round traffic is approximated by Maps traffic data within the most recent 2 months.  

\begin{figure}[!t]
\centering
\includegraphics[width=2.5in]{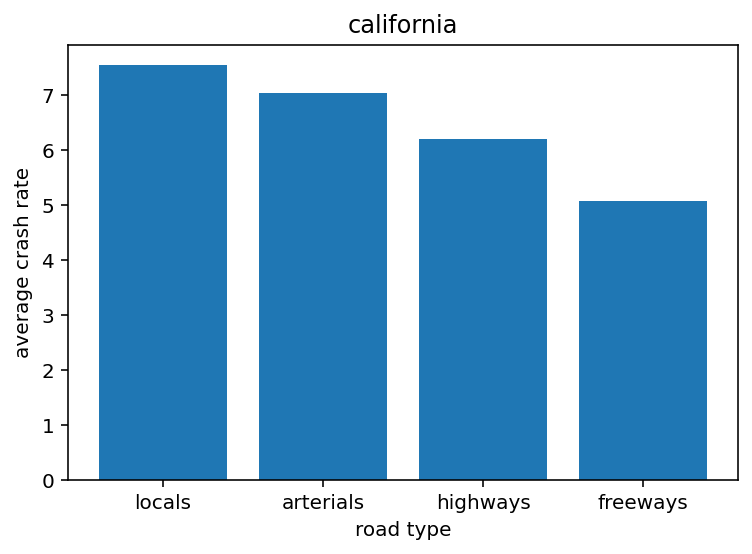}
\includegraphics[width=2.5in]{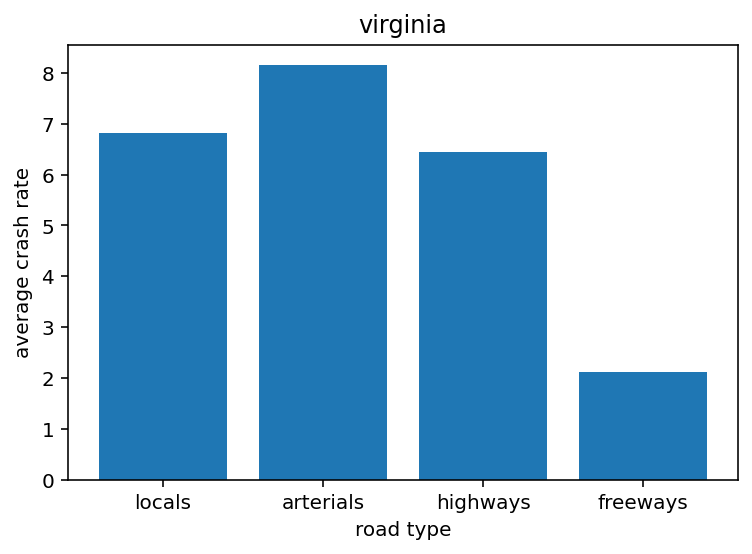}
\caption{Mean crash rate for different road categories for California (top plot) and Virginia (bottom plot)}
\label{fig_3}
\end{figure}

Figure~\ref{fig_3} compares the crash rates between different road categories. Local and arterial roads have slightly higher crash rates than freeways. The top (respectively, bottom) plot correspond to California (resp. Virgina) data. 

\begin{figure}[!t]
\centering
\includegraphics[width=2.5in]{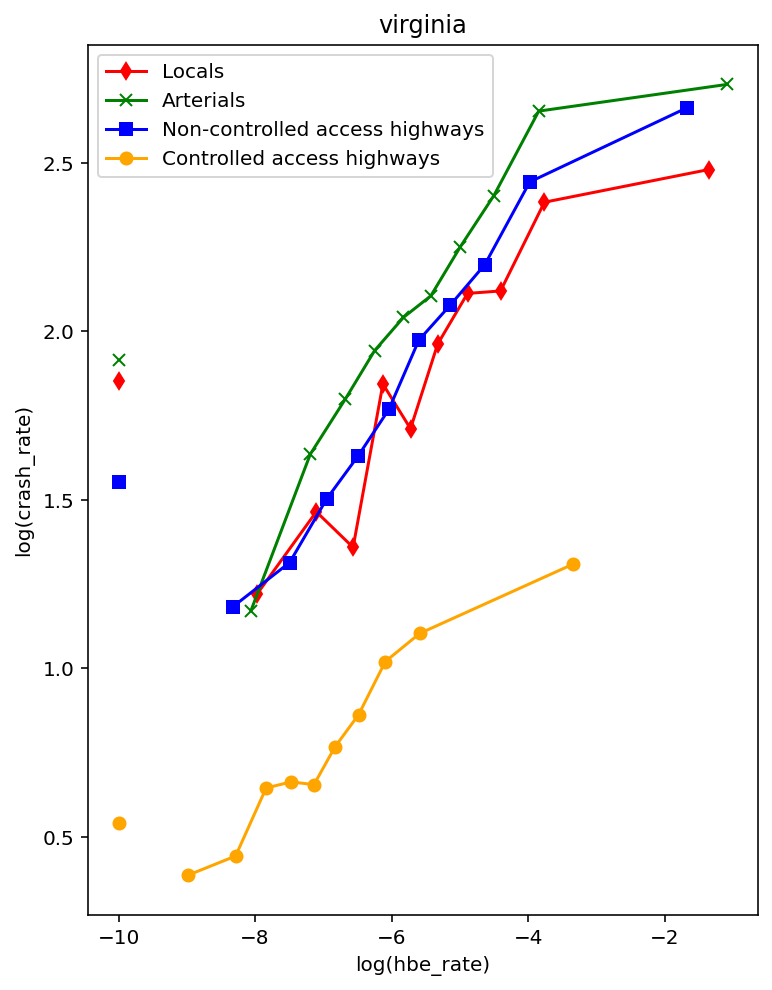}
\includegraphics[width=2.5in]{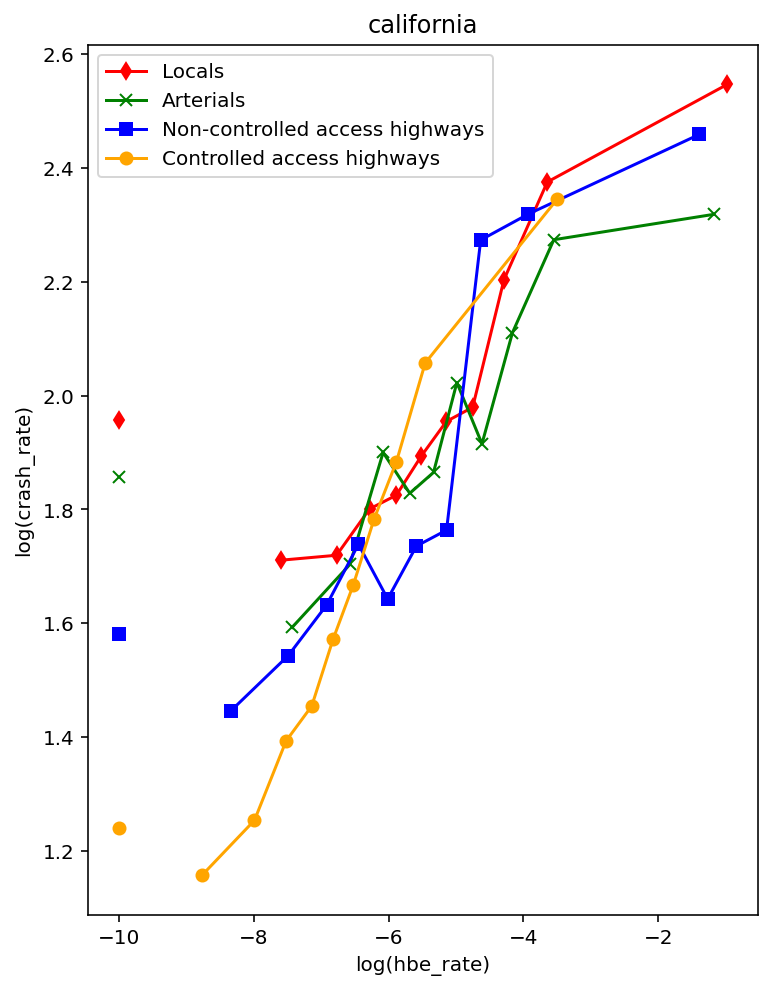}
\caption{Relationship between HBE rate and crash rate by road types for California (top plot) and Virginia (bottom plot)}
\label{fig_5}
\end{figure}

Figure~\ref{fig_5} shows the correlation between hbe rate and crash rate. The top (resp. bottom) plot corresponds to California (resp. Virigina) data. For each plot, there is one line per road class. For each line, road segments are binned by hbe rate into 10 equal-sized based on rank. For all segments that fall within a given hbe rate, we compute their mean crash rate. Note that the plots have logarithmic scales for both axes. Segments with a zero hbe rate are binned into the first bin, which is displayed, for reference, at $x=-10$.
These plots  clearly illustrate  a positive and linear correlation, on a logarithmic scale, between hbe rates and crash rates. 
When HBE rate is as low as 0, the crash rate is slightly higher than the first bin of a positive hbe rate. This is because for hbe rate = 0 we usually do not have enough traffic to observe a hard braking event during the time window. But for segments with hbe rate=0, their crash rate is overall much lower than those segments with high hbe rates, so are less concerned.

\subsection{Model outputs}
Poisson and Negative Binomial (NB) models are implemented for both Virginia and California. Since the Poisson model exhibited overdispersion, a common issue in crash modeling, the NB GLMs are used for formal statistical inference. The model incorporates multiple predictors, including crash rate, road type, number of lanes, presence of a ramp on the road, change in the number of lanes, and cumulative turning angle of the segment.  

As presented in Table II, the majority of predictors are statistically significantly associated with crash risk  ($\text{p-value} < 0.01$).  
To interpret the coefficients of the road type factors, note that controlled access highway road type, which includes interstate highways and expressways, is used as the reference. Roady type coefficents with a value below (resp. above) 1, indicate that the crash rate of that road type is lower (resp. higher) than that of controlled access highways.
For Virginia, all road type coefficients are above 1. This indicates that
the access-controlled highways show the lowest crash rate, followed by local and non-controlled access highways.  Arterials show the highest crash rate. For California, local roads and arterials show the highest crash rate.  Such discrepancy could be caused by multiple factors, such as local traffic flow, infrastructure, traffic management, and driver characteristics.  For both rates, the coefficient of the "segment contains ramps" factor is positive, indicating that the  crash rate increases on ramps. This  could be due to multiple risk factors like mandatory merging, navigating through the weaving section, and increased lane-changing and vehicle-to-vehicle interactions.    	
\begin{table*}[ht]
\centering
\begin{tabular}{|p{8cm}|c|c|c|c|}
\hline
& \multicolumn{2}{c|}{Virginia} & \multicolumn{2}{c|}{California} \\
\hline
& Estimate & Std. Error & Estimate & Std. Error \\
\hline
(Intercept) & $-0.81^{***}$ & 0.04 & $0.65^{***}$ & 0.02 \\
\hline
hbe\_rate & $0.23^{***}$ & 0.02 & $0.02^{***}$ & 0.007 \\
\hline
Locals vs controlled access highway & $1.22^{***}$ & 0.035 & $0.35^{***}$ & 0.02 \\
\hline
Arterials vs controlled access highway & $1.41^{***}$ & 0.03 & $0.35^{***}$ & 0.02 \\
\hline
Non-controlled access highways vs controlled access highway & $1.08^{***}$ & 0.03 & $0.24^{***}$ & 0.02 \\
\hline
Number of Drivable lanes & $0.35^{***}$ & 0.02 & $0.04^{***}$ & 0.005 \\
\hline
Segment contains ramp & $0.52^{***}$ & 0.05 & $1.24^{***}$ & 0.02 \\
\hline
Number of lanes changes in segment & $0.07^{***}$ & 0.01 & $-0.09^{***}$ & 0.007 \\
\hline
Cumulative turning angles & $-0.001^{***}$ & 0.0003 & $0.0002^{}$ & 0.0001 \\
\hline
\end{tabular}

\caption{ Negative Binomial regression results for crash rate versus log HBE rate.}
\label{tab:regression}
Sig Lv. 0 ‘***’ 0.001 ‘**’ 0.01 ‘*’ 0.05 ‘.’ 0.05 ‘ ’
\end{table*}

The HBE rate, which measures the frequency of hard braking events per mile driven, constitutes the central predictor of interest in this analysis. The results clearly indicate that the HBE-crash correlation is statistically significant across both the Virginia and California datasets. Specifically, the positive parameter estimates calculated for the HBE rate in both states demonstrate a strong and direct relationship: road segments exhibiting a higher frequency of HBEs are concurrently associated with a significantly elevated crash rate. This critical finding provides compelling empirical validation that the HBE metric is a reliable and substantially correlated proxy for crash risk at the individual road segment level. The correlation confirms that vehicles requiring frequent hard braking are likely operating in environments (road segments) that inherently pose greater risk for accidents.
It should be noted that the regression coefficients vary significantly across different states. This geographical disparity in coefficients may be attributed to a multitude of factors, including variations in local traffic conditions, differences in roadway infrastructure, and distinct local driving behaviors. It is crucial to account for these state-specific heterogeneities when interpreting the results.

\section{Case Study}


\begin{figure}[!t]
\centering
\includegraphics[width=2.5in]{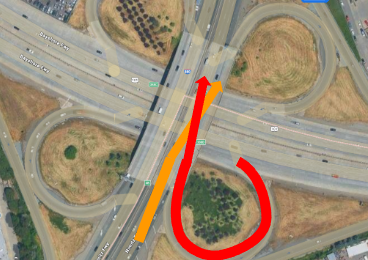}
\caption{CA freeway with highest crash rate}
\label{fig_7}
\end{figure}
\begin{figure}[!t]
\centering
\includegraphics[width=2.5in]{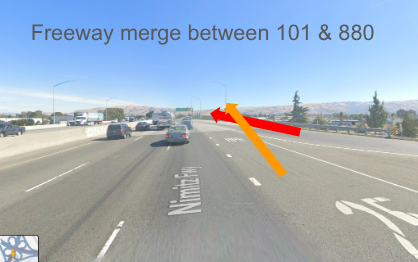}
\caption{Streetview of the merge location between freeways US-101 \& US-880}
\label{fig_8}
\end{figure}

Figures~\ref{fig_7} and \ref{fig_8} show images of the freeway in California with the high crash rate with 80+ crashes in the last 10 years. It consists of  two freeway (US-101) ramps, one on-ramp that merges traffic from another freeway (US-880), and one off-ramp that vehicles are switching from US-101 to US-880. Both freeways are the busiest freeways on San Francisco bay area, with large amount of commute traffic every day. Since the ramps connect two freeways, the driving speed is expected to be faster than ramps that connect local roads and freeways. As a result, large amount of on-ramp, off-ramp and through traffic cross each other with high speed, leading to lots of accelerations, decelerations by lane changes, within a very short distance (100 meters) between the two ramps, and can be extremely exhaustive and stressful to drivers. Our data shows that the HBE rate here is 70+ times higher than the average HBE rate on California freeways and top 1\% in terms of HBE rate among all CA road segments.

\section{Discussion}

A consistent, network-level, driving safety metric can support a wide spectrum of safety measures from proactive safe routing to identifying crash hotspots for improvement. Hard braking events detected from vehicle speed and smartphone sensors provide a low-cost and globally scalable metric for safety assessment. Successful implementation of such metrics can benefit the entire driver population and improve traffic safety.

One key issue for crash surrogate measures, such as HBEs, is whether they can represent crash risk for the intended applications. An essential requirement is that the risk of HBEs should be significantly correlated with crash risk. Using two state-level crash databases California and Virginia, this study  confirm that at the road segment level, the observed HBE rates are significantly associated with police-reported crash rates. Road segments with a high HBE rate tend to have a high crash rate, after adjustment for known factors such as road class, presence of ramp,  and lane number change. 


There are a few limitations worth recognizing. First, the HBE data and crash data are not collected in the same preiord, the HBE are collected from recent month while crashes are from several previous years. This temporal mismatch limits the relationship to factors that are relatively stable over a extended period of time, such as infrastracture characteristics and local driving behavior. 
Second, investigating mechanisms to spatially cluster homogenous segments such as to mitigate the data sparsity issue is of interest.

As empirically demonstrated in this study, Hard Braking Events  exhibit a statistically significant correlation with actual crash risk at the road segment level. The high spatio-temporal density of HBE data facilitates the timely identification of high-risk road segments, thereby circumventing the reliance on inherently lagging and sparse collision records. This research highlights  the potential of connected vehicle data to comprehensively assess traffic safety with enhanced spatial and temporal granularity across the entire road network.

\vfill

\end{document}